\documentclass[useAMS,usenatbib,usegraphicx]{mn2e}

\usepackage{bm}
\usepackage{amsmath}
\usepackage{amssymb}

\title[Discovery of the Aquarius~2 dwarf] {At the survey
  limits: discovery of the Aquarius~2 dwarf galaxy in the VST ATLAS and
  the SDSS data\thanks{Based on data products from observations made
    with ESO Telescopes at the La Silla Paranal Observatory under
    public survey programme ID programme 177.A-3011(A,B,C)}}

\author[G.~Torrealba et al.]  {G.~Torrealba$^1$, S.E.~Koposov$^1$,
  V.~Belokurov$^1$, M. Irwin$^1$, M. Collins$^{2,3}$,
  M. Spencer$^4$,\newauthor R. Ibata$^5$, M. Mateo$^4$, A. Bonaca$^3$
  and P. Jethwa$^1$ \\ $^1$Institute of Astronomy, Madingley Rd,
  Cambridge, CB3 0HA\\ $^2$Department of Physics, Faculty of
  Engineering and Physical Sciences, University of Surrey, Guildford,
  GU2 7XH\\ $^3$ Astronomy Department, Yale University, New Haven, CT
  06511 \\ $^4$University of Michigan, Department of Astronomy, 1085
  S. University, Ann Arbor, MI 48109\\ $^5$ Observatoire Astronomique
  de Strasbourg, 11 rue de l'Universit\'e, 67000 Strasbourg}
\begin{document}

\date{9th May 2016}

\pagerange{\pageref{firstpage}--\pageref{lastpage}} \pubyear{2015}

\maketitle

\label{firstpage}

\begin{abstract}

We announce the discovery of the Aquarius~2 dwarf galaxy, a new
distant satellite of the Milky Way, detected on the fringes of the VST
ATLAS and the SDSS surveys. The object was originally identified as an
overdensity of Red Giant Branch stars, but chosen for subsequent
follow-up based on the presence of a strong Blue Horizontal Branch,
which was also used to measure its distance of $\sim 110$ kpc. Using
deeper imaging from the IMACS camera on the 6.5m Baade and
spectroscopy with DEIMOS on Keck, we measured the satellite's
half-light radius $5.1\pm 0.8$ arcmin, or $\sim 160$ pc at this
distance, and its stellar velocity dispersion of $5.4^{+3.4}_{-0.9}$
km s$^{-1}$. With $\mu=30.2$ mag arcsec$^{-2}$ and $M_V=-4.36$, the new
satellite lies close to two important detection limits: one in surface
brightness; and one in luminosity at a given distance, thereby making
Aquarius~2 one of the hardest dwarfs to find.

\end{abstract}


\section{Introduction}\label{sec:INTRO}

The unprecedented combination of depth, coverage and stability of the
imaging data provided by the Sloan Digital Sky Survey \citep[SDSS, see
  e.g.][]{sdssdr5,sdssdr7,sdssdr8} has helped us to find what had gone
missing for a while: a large population of low-mass dwarf galaxies
\citep[see e.g.][]{Klypin1999,Moore1999,Koposov2008,
  Tollerud2008}. The bulk of the SDSS satellite discoveries happened
in the last decade \citep[see
  e.g.][]{Willman2010,Belokurov2013}. However, most recently, with the
announcement of the Pegasus dwarf, \citet{pegasus} have demonstrated
that the SDSS barrel has not been scraped clean yet.

The question, of course, is not whether the SDSS data contains clues
to the locations of further yet unidentified satellites,
but whether the avalanche of false positives encountered at
low significance levels can be filtered efficiently. The SDSS data
alone is not sufficient to confirm the nature of barely detectable
candidate stellar over-densities. Thus, deeper and/or better image
quality follow-up observations are required. Until recently, the
community lacked an appropriate tool for a fast and a cost-effective
follow-up of halo sub-structures. With the advent of the
DECam camera \citep[][]{decam} on the 4m Blanco telescope at Cerro
Tololo in Chile, surveying the Milky Way halo has been
revitalized \citep[see e.g.][]{McMonigal2014,Kim2015a, Mackey2016}.

However, even before embarking on a follow-up campaign, the candidate
lists can be purged further using information at hand. Curiously, the
lower luminosity SDSS satellite galaxies with distances in excess of
100 kpc (Her, Leo~IV, Leo~V, Peg) all seem to possess a noticeable 
Blue Horizontal Branch (BHB). As these are all Ultra Faint Dwarfs (UFDs), 
their BHBs are not very well
populated, but even a small handful of bright (and thus not easily
mis-classified as galaxies) and blue stars stands out dramatically
over the Milky Way foreground. At high Galactic latitudes, and above
$g=21$, the BHB stars suffer minuscule contamination from other
stellar populations. Moreover, they are reliable standard candles
\citep[see e.g.][]{Deason2011,sgrprecession}. Therefore, BHBs can be
used as a litmus test of a satellite's
presence. Typically, the ranking of the candidate is bumped up if
several BHBs (at a comparable distance) are detected in the vicinity
of an over-density of Main Sequence (MS) and/or Red Giant Branch (RGB) stars. 
An extreme and cunning version of this idea has recently been tested by
\citet{Sesar2014} who looked for evidence of stellar over-densities
around individual RR Lyrae stars, i.e. the pulsating sub-population of
BHBs.

As larger portions of the sky are surveyed and the census of
Galactic dwarfs is upgraded, previously unrecorded details of the
population spatial distribution start to emerge. While many of the
dwarfs lie close to the Magellanic plane \citep[see
  e.g.][]{LyndenBell1995,Jethwa2016}, ample examples now exist of
satellites not associated with the Clouds' infall. As a matter of
fact, some of these show hints of having been accreted onto the Milky
Way with a group of their own.  For example, claims have been made
that Segue~2 was once part of the bigger system that produced the
Tri-And structure and the Tri/Psc stellar stream \citep[see
  e.g.][]{Belokurov2009, DeasonTriAnd}. Moreover, Boo~II and Sgr~II
may be linked to the disrupting Sgr dwarf \citep[see e.g.][]{Koch2009,
  Laevens2015}. Most recently, \citet{Torrealba2016} presented
evidence for a possible connection between Crater, Crater~2, Leo~IV and Leo~V.

Interestingly, the satellite Galacto-centric distance distribution
appears to be at odds with the sub-halo radial density profiles
gleaned from the cosmological zoom-in simulations \citep[see
  e.g.][]{Springel2008}. According to Figure 6 of \citet{Jethwa2016},
compared to simulations, the current Milky Way satellites tend to
concentrate in excess at distances less than 100 kpc. This is
surprising, given the fact that the simulation sub-halo distribution
itself is likely biased towards higher densities at lower radii. This
is because the effect of the Galactic disk is typically not included
in simulations such as that of \citet{Springel2008}. As
\citet{Donghia2010} demonstrate, the baryonic disk will act to destroy
sub-halos with peri-centres within a few disk scale lengths. Hence,
the presence of the disk will flatten the sub-halo radial density
profile as compared to Dark Matter (DM) only simulations. There could
be several explanations of the measured Galactic dwarf satellite
excess at $R<100$ kpc. This can either be interpreted as a sign of the
accretion of a massive galaxy group with its own large entourage of
satellites. Alternatively, the flattening of the cumulative radial
density distribution beyond 100 kpc might not be the evidence for a
dwarf excess at smaller distances, but rather the consequence of a
further selection bias, causing a drastic drop in satellite detection
efficiency at higher distances.

In this paper we present the discovery of a new dwarf galaxy detected
in the SDSS and the VST ATLAS data. The previously unknown object in
the constellation of Aquarius was identified using a combination of
BHBs and MS/RGB stars as tracers. We have obtained deeper follow-up
imaging of the satellite with IMACS on Magellan as well as
spectroscopy with DEIMOS on Keck. The follow-up data has enabled a
robust measurement of the size of Aquarius~2 (hereafter Aqu~2) and 
velocity dispersion. Given
the substantial size and the enormous mass-to-light ratio returned by
our analysis, the satellite is most likely a dwarf galaxy. This paper
is organised as follows. Section~\ref{sec:Discovery} outlines the
discovery of the dwarf. Section~\ref{sec:FollowupImage} gives the
details of the follow-up imaging and the subsequent measurement of the
structural parameters of Aqu~2. Section~\ref{sec:spectrofollowup} deals
with the spectroscopic analysis.

\section{Discovery of Aqu~2}\label{sec:Discovery}

\begin{table}
    \caption{Properties of Aqu~2}
    \centering
    \label{tab:Properties}
    \begin{tabular}{@{}lrl}
        \hline
        Property               & Value                         & Unit\\
        \hline
        $\alpha({\rm J2000})$   & $338.4813 \pm 0.005$          & deg \\
        $\delta({\rm J2000})$   & $-9.3274 \pm 0.005$           & deg  \\
        Galactic $l$            & $55.108$                      & deg \\
        Galactic $b$            & $-53.008$                      & deg \\
        $(m-M)$                & $20.16 \pm 0.07$              & mag\\
        $D_\odot$              & $107.9 \pm 3.3$               & kpc\\
        $r_{h}$                & $5.1 \pm 0.8$                 & arcmin\\
        $r_{h}$                & $159 \pm 24$                  & pc\\
        1$-$b/a                & $0.39 \pm 0.09$               & \\
        PA                     & $121 \pm 9$                   & deg\\
        $M_V$                  & $-4.36 \pm 0.14$                & mag\\
        $V_{helio}$            & $-71.1 \pm 2.5$               & km/s\\
        $V_{\rm GSR}$            & $49$               & km/s\\
        $[$Fe/H$]$             & $-2.3\pm0.5$                  & dex\\
        $\sigma_v$             & $5.4^{+3.4}_{-0.9}$           & km/s\\
        $\rm{Mass}(<r_h)$      & $2.7^{+6.6}_{-0.5}\times10^6$ & $M_\odot$\\
        $M/L_V$                & $1330^{+3242}_{-227}$         & $M_\odot/L_\odot$\\
        \hline
    \end{tabular}
\end{table}

\subsection{VST ATLAS}

ATLAS \citep{Shanks2015} is one of the three public ESO surveys
currently being carried out using the 2.6\,m VLT Survey Telescope
(VST) at the Paranal observatory in Chile. The VST is equipped with
a 16k $\times$ 16k pixel CCD camera OmegaCAM, which provides a
1-degree field of view sampled at 0.\arcsec21 per
pixel. ATLAS aims to survey 4500 square degrees of the Southern
celestial hemisphere in 5 photometric bands, $ugriz$, with depths
comparable to the Sloan Digital Sky Survey (SDSS). The median limiting
magnitudes, corresponding to the $5\sigma$ source detection limits,
are approximately 21.99, 23.14, 22.67, 21.99, 20.87 for each of the
$ugriz$ bands, respectively. Image reduction and initial catalog
generation are performed by the Cambridge Astronomical Survey Unit
(CASU) \citep[see][for details]{Koposov2014}. The band-merging and
selection of primary sources were performed as separate steps using a
local SQL database. To improve the uniformity of the photometric
calibration of the survey, on top of the nightly zero-points measured
relative to APASS survey, we also applied an additional global
calibration step \citep[a.k.a. uber-calibration;][]{Padma08}. In this
work, we use the photometric catalogs provided by CASU, which include
the entirety of the VST ATLAS data taken up to September 2015 covering
$\sim$ 4500 square degrees in at least one band, and with $\sim$3500
square degrees having both $g$ and $r$ band observations. In the
analysis that follows we correct ATLAS photometry for the effects of
Galactic dust extinction using the \citet{SFD} maps and the extinction
coefficients from \citet{Schlafly2011}.

\begin{figure*}
    \includegraphics[width=\textwidth]{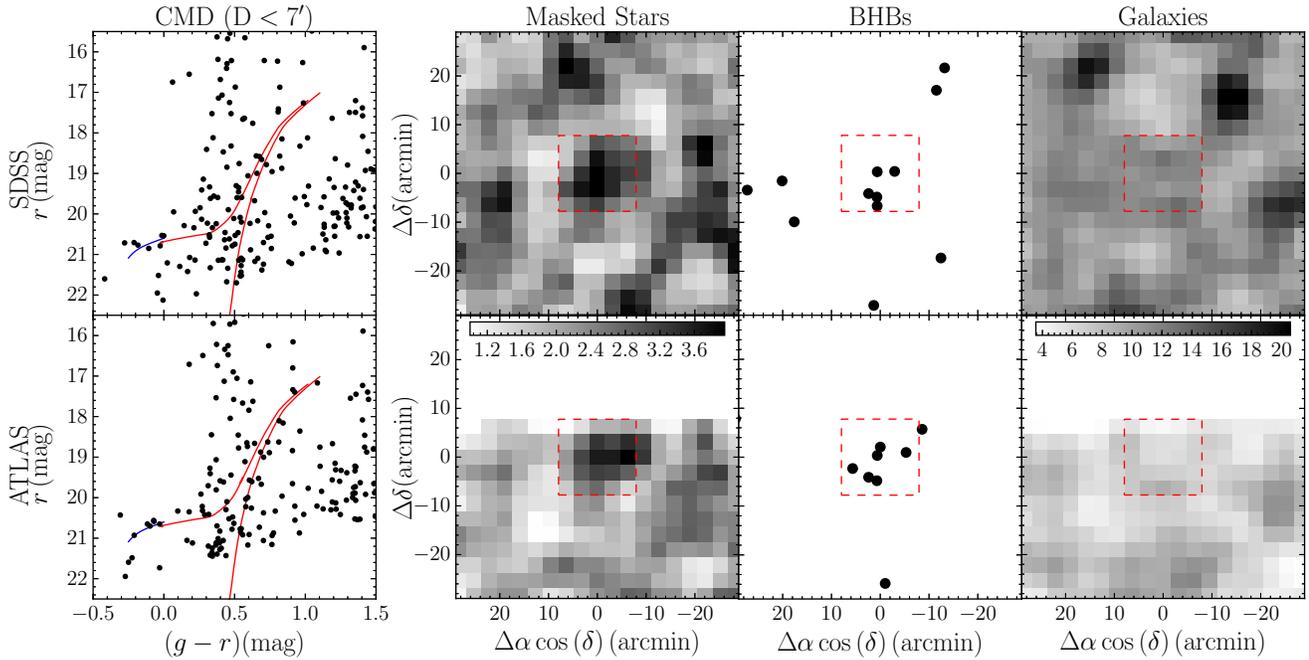}
    \caption{Aqu~2 as seen by ATLAS and SDSS. The panels on the top
    shows data from SDSS DR9, and the panels in the bottom show data from
    ATLAS. The leftmost panels show the CMD of the stars within 7\arcmin\ 
    of the center of Aqu~2. The red line shows a PARSEC isochrone with
    [Fe/H]=-1.8, age $12$\,Gyr and m$-$M=20.16, and the blue line shows the
    outline of the blue horizontal branch from \citet{Deason2011}. In both 
    SDSS and ATLAS
    CMDs there is an
    obvious overpopulation of BHB stars, but only in ATLAS are there hints
    of an RGB. In the middle left panel we show a density map (in units of objects per pixel) of stars inside the isochrone mask shown in Figure \ref{fig:FollowupCMDMarked}, 
    and in the middle right panel we show the BHBs
    present in the field. In both tracers there is a visible overdensity.  
    Finally, the rightmost panel shows the density map of galaxies,
    demonstrating that there is no obvious overdensity of galaxies 
    at the location of Aqu~2.} \label{fig:OriginalDiscovery}
\end{figure*}

\subsection{Discovery}

We discovered Aqu~2 by sifting through the ATLAS data armed with a
version of the stellar overdensity detection algorithm \citep[see
  e.g.][]{Irwin1994,Koposov2008, Walsh2009, Koposov2015,
  Torrealba2016}. Briefly, the algorithm starts by filtering stars
using an isochrone mask at a given age, metallicity and distance. The
local density of stars is then measured and compared to the density at
larger scales, i.e. the Galactic background/foreground. In practice,
this is done by convolving the masked stellar number count
distribution with a ``Mexican hat'' kernel: a difference between a
narrow inner kernel (for the local density estimation) and a wide
outer kernel (to gauge the background density). In our implementation,
both kernels are two-dimensional Gaussians and the significance of the
detection at each pixel is calculated by comparing the result of the
convolution with the expected variance.

In the current implementation, we run the algorithm with two sets of
isochrone masks: one that selects both MS and RGB stars, and a second
one to pick out the likely BHB stars. The first mask is based on the
most recent PARSEC evolutionary models \citep{Bressan2012}, from
which, for simplicity, only old (12 Gyr) and metal poor
($[$Fe/H$]=-2$) population is chosen. The second mask is constructed
from the BHB absolute magnitude ridgeline as a function of the $g-r$
color from \citet{Deason2011}. In both cases the masks widths are
defined by the observed photometric errors above the minimum width of
0.1 magnitudes, in the case of PARSEC isochrones, and 0.2 magnitudes
in the case of the BHBs (see Figure \ref{fig:FollowupCMDMarked}).

We applied the above search method to the ATLAS data using a grid of
inner kernel sizes (from 1\arcmin\ to 10\arcmin), and a grid of
distance moduli ($15<m-M<23$) for both masks. The outer kernel size
was fixed to 60\arcmin. As a result, two candidate lists were created,
one for MS/RGB stars and one for BHBs. These were then cross-matched
to create a single list of objects that were identified using both
sets of tracers. In this combined list, Aqu~2 stood out as the
only object of unknown nature, detected with a significance of
$5.6\sigma$ in MS/RGB, and $4.9\sigma$ in BHBs. Aqu~2 also
happens to be located within the footprint of the SDSS DR9, although
it only had a significance of $4\sigma$ in MS/RGB stars, but a
$5.5\sigma$ detection in BHBs. Figure \ref{fig:OriginalDiscovery}
shows Aqu~2 as detected in both SDSS (top row) and ATLAS (bottom
row). Overall, there is a strong evidence for a promising satellite
candidate: in ATLAS, the MS/RGB stellar density map shows a clear
overdensity (second panel, bottom row), which is not matched by any
obvious galaxy clumping (fourth panel, bottom row). In the
color-magnitude diagram (CMD, left panel) a hint of an RGB at $m-M\sim
20.16$ can be discerned between $19 < r< 22$. These are complemented
by a handful of BHBs at a similar distance, which cluster tightly
around the satellite location (third panel). Traces of a stellar
overdensity can also be seen in the SDSS DR9 data (top row). While the
SDSS RGB is not as prominent, the BHBs are readily identifiable, in
perfect agreement with the significance values reported by the
systematic search.

\begin{figure*}
    \includegraphics[width=\textwidth]{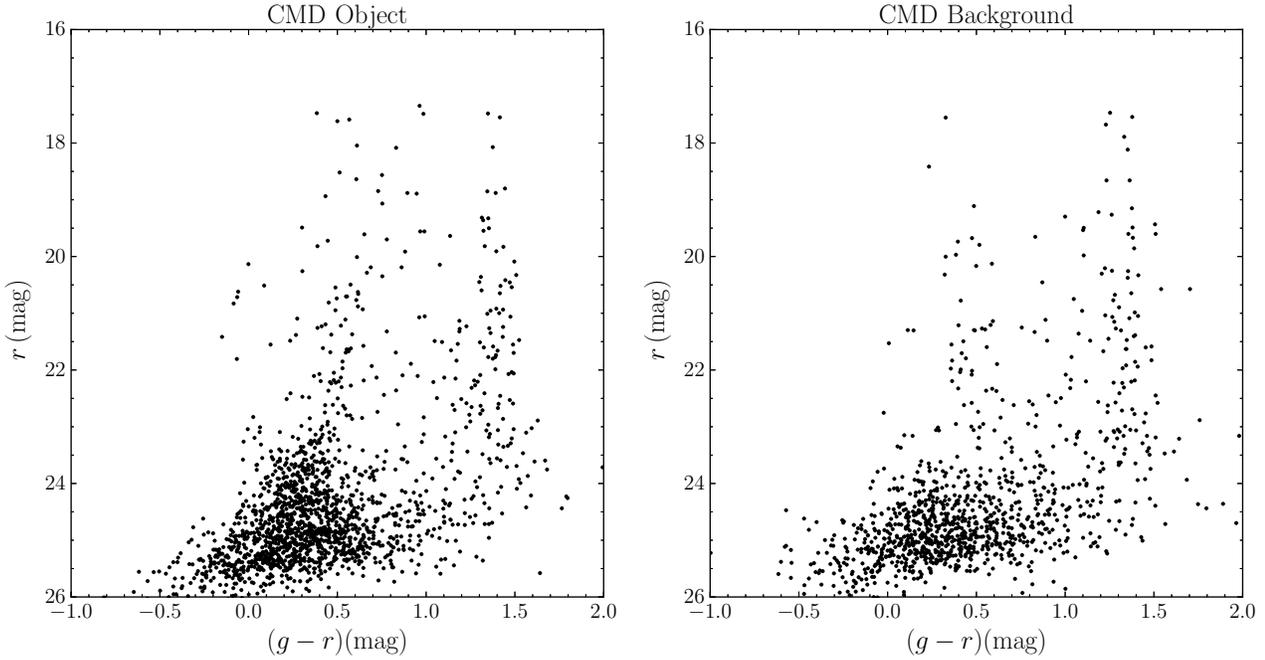}
\caption{Color-magnitude diagram of Aqu~2 based the IMACS follow-up
  imaging. The left panel shows stars inside the half-light radius
  $r_h$, while the right panel shows stars from an equal area outside 
  7\arcmin\ from the center. The CMD of the central areas of Aqu~2
  demonstrates very clear evidence of the main sequence turn-off at r
  $\sim$ 24, the red giant branch and the horizontal branch at r
  $\sim$ 21.  These features are all consistent with the stellar
  population at a distance modulus of m$-$M$\sim20.16$
  ($\sim108$\,kpc) and match well with the features visible in the VST
  and SDSS data.} \label{fig:followupCMD}
\end{figure*}

\section{Photometric follow-up}\label{sec:FollowupImage}

\subsection{IMACS imaging}

To confirm the nature of the candidate overdensity described above,
deeper photometric data was obtained using the 6.5\,m Walter Baade
telescope at the Las Campanas observatory in Chile. The images were
taken on 19~Aug 2015 using the f/4 mode of the Inamori-Magellan Areal
Camera \& Spectrograph (IMACS) which provides a
15.4$\arcmin\times$15.4$\arcmin$ field of view with a mosaic of 8 2k$\times$4k 
CCDs \citep{Dressler2011}.  We employed 2x2 binning which gives a
pixel scale of $\sim$ 0.22$\arcsec$/pixel. A total of 4 exposures centered
on Aqu~2 were taken in each of the $g$ and $r$ filters giving a total
exposure time of 19 minutes in $g$ and 23.5 minutes
in $r$. After standard reduction steps, the images were
astrometrically calibrated using {\it astrometry.net} software
\citep{Lang2010} and stacked using the SWARP software \citep{Bertin2002}.

\subsection{Catalog Generation}

The IMACS object catalogs were created by performing photometry on the
images using SExtractor/PSFex \citep{Bertin1996} as described in
\citet{Koposov2015}. First, an initial SExtractor pass was done to
create the source list required by PSFex. Then, PSFex is run 
to extract the point spread function (PSF) shape, which
is then used in the final SExtractor pass. This procedure delivers
estimates of the model and the PSF magnitudes for each object, as well
as estimates of the \texttt{SPREAD\_MODEL}
and \texttt{SPREADERR\_MODEL} parameters. 
The final
catalog is assembled by merging the g and r bands within a cross-match
distance of 0.7$\arcsec$, only selecting
objects with measurements in both bands. We calibrated the
instrumental magnitudes by cross-matching the catalog with the SDSS.
The resulting zero-point is measured with a photometric precision of
$\sim0.1$ mag in both bands. Finally, likely stars are separated from
likely galaxies by selecting all objects with
$|\texttt{SPREAD\_MODEL}|<0.005+\texttt{SPREADERR\_MODEL}$ \citep[see
e.g.][]{Desai2012,Koposov2015}.

Figure \ref{fig:followupCMD} shows the CMD of the area around Aqu~2
constructed using the merged stellar catalogs from the IMACS data.
Only stars within the half light radius of Aqu~2 (see
Section~\ref{sec:Structural} for details on its measurement) are
displayed in the left panel, while the stars of an equal area outside
the central region are given in the right panel for comparison. The
left panel of the Figure leaves very little doubt as to the nature of
the candidate: the familiar CMD features of a co-eval and co-distant
old stellar population, a very prominent MS turn-off, the strong RGB
and a clumpy BHB are all conspicuous. Note that while the CMD shown in
the right panel contains mostly Galactic foreground, there are hints
of the RGB and MSTO of Aqu~2 here too, which is consistent with the
object extending beyond its half light radius (see
Section~\ref{sec:Structural} for further discussion).  Figure
\ref{fig:FollowupCMDMarked} shows the same CMD as in the left panel of
Figure \ref{fig:followupCMD} but with various stellar sub-populations
highlighted. To guide the eye, the PARSEC isochrone with
[Fe/H]=$-1.8$, age of 12 Gyr, and offset to $m-M= 20.16$ is
over-plotted in red. This isochrone is used to create the CMD
  mask to select stars belonging to the satellite. Thus, the age and
  the metallicity of the isochrone were chosen to maximize the number
  of member stars within the mask. Orange, red, and blue filled
circles mark stars identified as foreground, Aqu~2 RGB and Aqu~2 BHB
stars respectively based on follow-up spectroscopy (see
Section~\ref{sec:spectrofollowup} for details). The red dashed line
shows the isochrone mask used to select the likely MSTO and RGB
members of the satellite, and the region in blue is the area where
probable BHB member stars lie. The shape and the position of the blue
area is based on the BHB ridge- line given in
\citet{Deason2011}. These likely BHB member stars are also used to
measure the distance modulus of Aqu~2 as $m-M=20.16\pm 0.07$, which
corresponds to a heliocentric distance of $108 \pm 3$ kpc.

\subsection{Structural Parameters}\label{sec:Structural}

\begin{figure}
    \includegraphics[width=0.98\columnwidth]{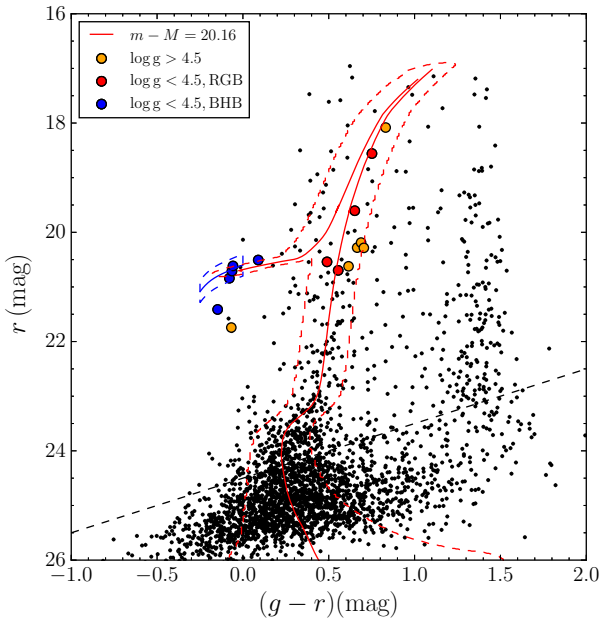}
    \caption{Color magnitude diagram of the central 7\arcmin\ of Aqu~2. 
    The red line is a PARSEC isochrone with
    [Fe/H]=-1.8, age $12$ Gyr and m$-$M=20.16. 
    The blue dashed contour
    marks the selection area of BHB stars (with 5 stars) used to constrain the
    distance to Aqu~2  and the dashed red line delineates the isochrone mask.  
    The dashed black line
    corresponds to $g=24.5$, which is the limiting magnitude that we 
    use when determining structural parameters of Aqu~2.  Large
    filled circles mark stars with measured spectra, in orange we show 
    likely dwarf stars with $\log({\rm{g})}>4$ (representing contamination), 
    and in blue and red we show giants with $\log({\rm{g})}<4$ (likely BHB and
    RGB stars respectively).
   }
    \label{fig:FollowupCMDMarked}
\end{figure}

The structural parameters of Aqu~2 are determined by modelling the
distribution of the likely member stars based on their colors and magnitudes 
(i.e. those located inside the isochrone mask outlined with the red 
dashed line in Figure~\ref{fig:FollowupCMDMarked}) 
in the IMACS field of view \citep[see
  e.g.,][for a similar
  approach]{Martin2008,Koposov2015,Torrealba2016}. To reduce the
contamination from galaxies misclassified as stars, we
only use stars brighter than $g=24.5$. The spatial model for 
the stellar distribution consists  of a flat Galactic
background/foreground density and a 2-dimensional elliptical Plummer
profile \citep{Plummer1911}. The Plummer profile is defined as:

\begin{equation}
P_{obj}(x,y|\Theta)=\frac{1}{\pi a^2 \left(1-e\right)}\left(1+\frac{\tilde{r}^2}{a^2}\right)^ {-2},
\label{eq:plummer}
\end{equation}
\begin{figure} \includegraphics[width=0.98\columnwidth]{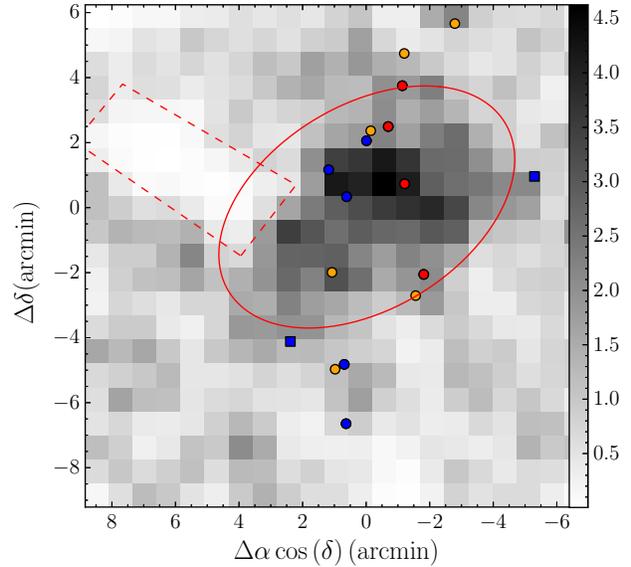}
    \caption{ Density maps of stars within the isochrone mask from the
      follow-up imaging.  The red solid line shows the elliptical
      half-light contour of the best fit model, and the red dashed
      line marks a region with the bright star that was excised from
      the analysis.  The orange, red and blue circles mark the
      locations of stars spectroscopically confirmed as foreground,
      RGB and BHB stars respectively.  Blue squares show BHB stars
      photometrically selected using the blue box of Figure
      \ref{fig:FollowupCMDMarked}, but without measured spectra.  }
    \label{fig:FollowupDM}
\end{figure}

\noindent where $x$ and $y$ are the on-sky coordinates which have been
projected to the plane tangential to the center of the object,
$\Theta$ is a shorthand for all model parameters, namely, the
elliptical radius $a$, the coordinates of the center $x_0$,
$y_0$, the positional angle of the major axis $\theta$, and the
ellipticity of the object $e$:
\begin{eqnarray}
\tilde{r}&=&\sqrt{\tilde{x}^2+\tilde{y}^2} \\
\tilde{x}&=&(\left(x-x_0\right)\,\cos\theta-\left(y-y_0\right)\sin\theta )/\left(1-e\right) \nonumber\\
\tilde{y}&=&\left(x-x_0\right)\sin\theta+\left(y-y_0\right)\cos\theta\nonumber.
\end{eqnarray}

\noindent The probability of observing a star at $x,y$ is then:

\begin{equation}
P(x,y|\Theta)=\frac{f}{A_{obj}(\Theta)}\,P_{obj}(x,y|\Theta)+(1-f)\,\frac{1}{A},
\label{eq:sampling_distr}
\end{equation}

\noindent with $A$ being the area of the data footprint, 
$f$ is the fraction of stars belonging to the object rather than foreground
and $A_{obj}(\Theta)$ is the integral of the Plummer model from
Eq.~\ref{eq:plummer} over the data footprint.

With Eq.~\ref{eq:sampling_distr} defining the distribution for
positions of stars,
we sample the posterior distribution of parameters of the model 
$P(\Theta|D)$ using MCMC. We use the affine invariant
ensemble sampler \citep{GoodmanWeare2010} implemented as an 
{\it emcee} Python module by \citet{ForemanMackey2013} with flat priors on all the
model parameters except $a$, in which we use the Jeffreys prior $P(a)\propto\frac{1}{a}$. 
The best-fit parameters were determined from the mode of the marginalized 1D
posterior distributions, with error-bars determined from the 16\% 84\%
percentiles of the posterior distribution.
The best-fit model returns a marginally elliptical Plummer 
profile with $e=0.4 \pm 0.1$ and an elliptical half-light radius of
$r_h=5\arcmin.1\pm0\arcmin.8$ corresponding to a physical size of $\sim$ 160\,pc.

\begin{figure}
    \includegraphics[width=0.98\columnwidth]{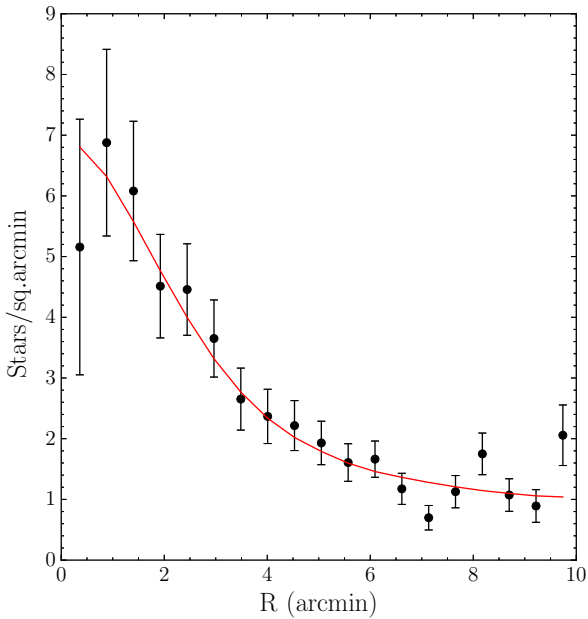}
    \caption{Radial number density profile of Aqu~2 as probed by the
    stars inside the isochrone mask shown in Figure \ref{fig:followupCMD}. 
    The red line shows the best fit model, which is formed by a flat
    background and a elliptical Plummer profile with $r_h=5\arcmin.1$ and
    $1-\rm{b/a}=0.39$} \label{fig:RadialProfile}
\end{figure}

The density map of isochrone selected stars is shown in Figure
\ref{fig:FollowupDM}. The dark elliptical blob in the center of the image is
Aqu~2. The red solid line gives the half-light contour of the
best-fit model while the red dashed line marks the region affected by a
bright star, that was masked out. Orange, red and blue filled circles give the
positions of the handful of stars with measured spectra as described
above. 
Figure~\ref{fig:RadialProfile} presents the 1D radial profile
of Aqu~2 together with the best-fit model (red line), which
clearly provides an adequate fit to the data in hand.

We estimate absolute luminosity of Aqu~2 by first counting the
number of observed Aqu~2 members. We do this by measuring the
fraction of member stars, $f$, in Eq.~\ref{eq:sampling_distr} using
a broad isochrone mask (with a limiting width of 0.2 magnitudes),
and with all the parameters of the model but $f$, fixed.  For this
calculation we use a limiting magnitude of $\sim$24.5 in both $g$
and $r$ bands, as above this limit the point source completeness of
our catalog is higher than 90\%. The fraction $f$ of the total
number of stars belonging to Aqu~2 that are brighter than 24.5 in
both $g$ and $r$ is measured to be $0.50 \pm 0.04$, which translates
into $282 \pm 24$ stars. Subsequently, we can now compute the
absolute magnitude of Aqu~2 by assuming a single stellar population
following a PARSEC isochrone with Chabrier initial mass function
\citep{Chabrier2003}. Using the Parsec isochrone with metallicity of
$-1.8$ with an age of 12 Gyr yields an estimation of the absolute
luminosity of $M_V = -4.27 \pm 0.1$ without including BHBs. Note
that the contribution from BHBs is added separately, owing to the
uncertainties in the modeling of HB morphologies \citep[see
  e.g.][]{Gratton2010}.  Assuming that Aqu~2 contains $\sim$ 7 BHB
stars, we derive their contribution to the luminosity of the object
$M_V(BHB)=-1.62 \pm 0.09$. Combined contribution from stars within
the isochrone mask and the BHBs added together gives the final
luminosity for Aqu~2 of $M_V=-4.36\pm0.14$.  We note that while the
metallicity of the isochrone used for calculating the luminosity of
the object is higher than the metallicity that we derived
spectroscopically $[Fe/H]=-2.3 \pm 0.5$ (see
Section~\ref{sec:spectrofollowup} for details), we have verified
that our estimate of the Aqu~2 luminosity is not sensitive to
changes of metallicity within 0.2 dex.


\section{Spectroscopic follow-up}\label{sec:spectrofollowup}

Aqu~2 was observed on the night of 11th September 2015, using the DEep
Imaging Multi-Object Spectrograph (DEIMOS) on the 10\,m Keck II
telescope. DEIMOS is a slit-based spectrograph, and one mask centered
on Aqu~2 was produced using the DEIMOS mask design software, {\texttt
  DSIMULATOR}, with VST ATLAS photometry described above. Stars were
prioritized for observation based on their position in the CMD of
Aqu~2, and proximity to the photometric centre of the object. Only
stars with $i-$band magnitude $<22$ were selected for observation, to
ensure that our final spectra had sufficient $S/N$ to determine a velocity ($S/N>3$ per pixel). Out 20 potential
member stars observed, 15 were successfully reduced. These 15 stars are identified as large filled circles on
Figures \ref{fig:FollowupCMDMarked} and \ref{fig:FollowupDM}.  The
Aqu~2 mask was observed with a central wavelength of 7800\AA, using
the medium resolution 1200 line/mm grating ($R\sim1.4$~\AA\ at our
central wavelength), and the OG550 filter. This gave a spectral
coverage of $\sim5600-9800$~\AA, isolating the region of the calcium
II triplet (Ca II) at $\lambda\sim8500$~\AA. The target was observed
for 1 hour, split into $3\times20$ minute integrations. The conditions
were ideal, with seeing ranging from $0.4-0.5^{\prime\prime}$.

We reduced the spectra using our standard DEIMOS pipeline, described
in \citet{Ibata2011} and \citet{COllins2013}. Briefly, the pipeline
identifies and removes cosmic rays, corrects for scattered light,
performs flat-fielding to correct for pixel-to-pixel variations,
corrects for illumination, slit function and fringing, wavelength
calibrates each pixel using arc-lamp exposures, performs a
two-dimensional sky subtraction, and finally extracts each
spectrum without resampling in a small spatial region around the
target. 

\subsection{Spectral modelling}

\begin{figure*} \includegraphics[width=\textwidth]{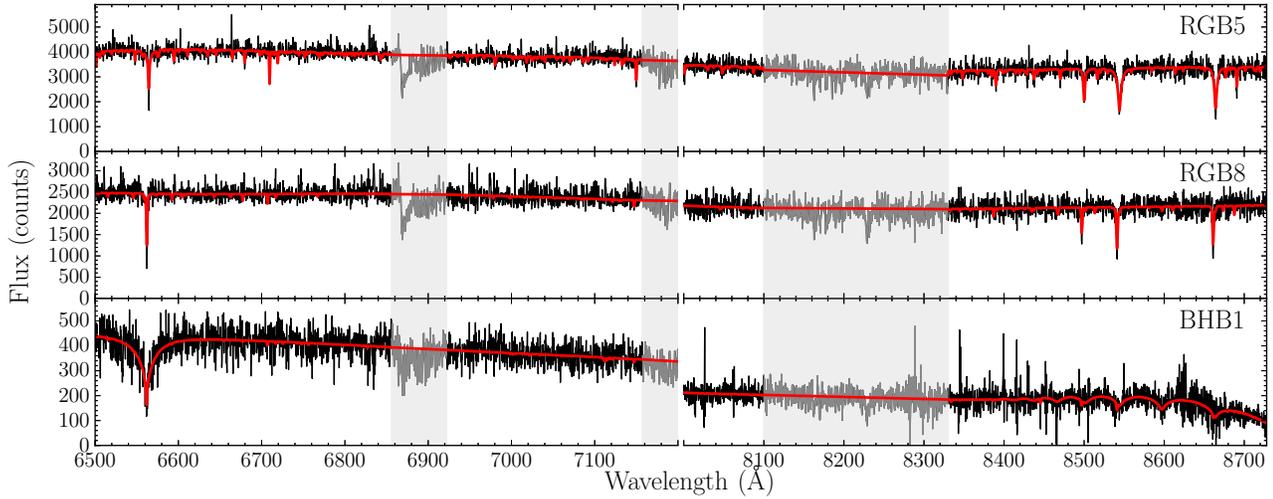}
    \caption{Example of the observed spectra (black) together with our
      best fit model (red) for a field star (top), an RGB star
      (middle), and a BHB star (bottom).  The grey-shaded regions mark
      regions affected by telluric absorption that were masked out
      during model fitting of the spectra.  } \label{fig:Specexample}
\end{figure*}

The spectroscopic properties of the stars observed were inferred using
a procedure similar to that described in \citet{Koposov2011,
  Koposov2015b}, i.e., via direct pixel-fitting of a suite of template
spectra to the observed spectrum. The template spectra were drawn from
the medium resolution PHOENIX library \citep{Husser2013}. The library
spectra are given at a resolution of $R=10000$ in a wavelength range
between $3000$\,\AA\ and $25000$\,\AA\ and cover a wide range of stellar
parameters: $2300$\,K$\leq\rm{T}_{\rm{eff}}\leq12000$\,K;
$0\leq\log{\rm{g}}\leq6$; $-4\leq[$Fe/H$]\leq1$; and
$-0.2\leq[\alpha$/Fe$]\leq1.2$. In practice, we pick values in a model
grid using a 100 K step size for $\rm{T}_{\rm{eff}}<7000$ and 200 K
for $\rm{T}_{\rm{eff}}>7000$, 0.5 dex in $\log{\rm{g}}$, 0.5 dex for
$[$Fe/H$]>-2$ and 1 dex for $[$Fe/H$]<-2$, and two discrete values for
$[\alpha$/Fe$]=\{0,0.4\}$. 
All  template spectra are degraded
from the initial resolution to the observed resolution of R=$5900$. 

Following \citet{Koposov2011,Koposov2015b}, in order to account for the
absence of flux calibration of the spectra the models are
represented as template spectra multiplied by a polynomial of 
wavelength of degree $N$ defined as:

\begin{equation}
P(\lambda)=\sum\limits_{j=0}^{N}a_j\lambda^j,
\end{equation}

\noindent where $a_j$ are the polynomial coefficients. Thus, the final
model spectrum is as follows:

\begin{equation}
M(\lambda,i,V)=P(\lambda)T_i\left(\lambda\left(1+\frac{V}{c}\right)\right),
\end{equation}

\noindent where $T_i$ is the downsampled $i$-th template.

The log-likelihood of each model is then defined as $-0.5\chi^2$, where
$\chi^2$ is calculated summing the squared standardized residuals
between the data and the model over all pixels in the spectrum:

\begin{equation}
\chi^2(i,V)=\sum\limits_k\left(\frac{O(\lambda_k)-M(\lambda_k,i,V)}{E(\lambda_k)}\right)^2,
\end{equation}

\noindent where $O(\lambda_k)$ is the observed spectra, $E(\lambda_k)$
the uncertainties, and $\lambda_k$ the wavelength at each
pixel. Examples of the three observed spectra, a field star, and an RGB and a BHB, both
members of Aqu~2, with their corresponding best-fit models are shown in
Figure \ref{fig:Specexample}.

For each star, the best-fit spectral model is obtained in two
steps. First, we test velocities between -700 km\,s$^{-1}$ and 700
km\,s$^{-1}$ with a resolution of 5 km\,s$^{-1}$ - while
simultaneously varying other model parameters - to zoom-in onto the
approximate heliocentric velocity solution. Next, we refine the
velocity grid resolution to 0.5 km\,s$^{-1}$ in the region around the
best-fit velocity obtained in the first pass. The final velocity and
the stellar atmosphere parameters
($T_{\rm{eff}},\log{\rm{g}},[$Fe/H$]$ and $[\alpha$/Fe$]$) are found
by marginalizing the likelihood over the nuisance parameters. We
choose to use flat priors on all parameters apart from temperature,
for which we use Gaussian priors based on the $g-r$ colour of the star.
The temperature prior is dictated by the colour-temperature
relation obtained for the VST-ATLAS stars with 
spectroscopic effective temperatures measured by the SEGUE Stellar Parameter
Pipeline \citep{Lee2008a,Lee2008b,Allende2008}. Finally, the preferred values
for the model parameters, and their associated uncertainties, are
measured from 1-D marginalized posterior probability distributions,
corresponding to the $16\%, 50\%$ and $84\%$ percentiles. These
are reported in Table~\ref{tab:SpectraTable}.

\begin{table*}
    \caption{Stellar parameters of spectroscopic targets}
    \centering
    \label{tab:SpectraTable}
    \begin{tabular}{@{}lrrrrrrrrrr}
        \hline
        Object & $\alpha$(J2000) & $\delta$(J2000) & g & r & $V_h$ & $T_{\rm eff}$ & $\log{\rm g}$ & $[$Fe/H$]$ &
$\chi^2$/dof & Member?\\
        &(deg)&(deg)&(mag)&(mag)&(km/s)&(K)&(dex)&(dex)\\
        \hline
        RGB1&$338.45496$&$-9.37256$&$21.24$&$20.62$&$-115.9\pm1.2 $&$4896_{-31 }^{+347}$&$-$&$-1.5_{-0.0}^{+0.4}$&2.21&N\\
        RGB2&$338.45067$&$-9.36172$&$21.03$&$20.54$&$-78.7 \pm2.0 $&$5090_{-21 }^{+46 }$&$3.3_{-0.4}^{+0.3}$&$-3.0_{-0.1}^{+0.0}$&2.16&Y\\
        RGB3&$338.46083$&$-9.31522$&$20.26$&$19.6 $&$-75.1 \pm1.0 $&$5093_{-10 }^{+5  }$&$2.4_{-0.1}^{+0.1}$&$-2.0_{-0.0}^{+0.0}$&4.57&Y\\
        RGB4&$338.49946$&$-9.36058$&$20.95$&$20.28$&$-131.4\pm1.5 $&$5233_{-136}^{+246}$&$4.6_{-0.4}^{+0.7}$&$-2.0_{-0.0}^{+0.4}$&2.3 &N\\
        RGB5&$338.47883$&$-9.28794$&$18.91$&$18.08$&$43.5  \pm0.3 $&$5098_{-3  }^{+1  }$&$5.0_{-0.0}^{+0.0}$&$-0.5_{-0.0}^{+0.0}$&9.62&N\\
        RGB6&$338.46208$&$-9.26497$&$21.25$&$20.7 $&$-77.7 \pm3.6 $&$4970_{-127}^{+257}$&$3.1_{-0.9}^{+0.9}$&$-2.1_{-0.3}^{+0.1}$&1.85&Y\\
        RGB7&$338.46104$&$-9.24833$&$20.88$&$20.19$&$-247.4\pm2.6 $&$5095_{-9  }^{+4  }$&$5.8_{-0.3}^{+0.2}$&$-1.0_{-0.1}^{+0.0}$&2.51&N\\
        RGB8&$338.46954$&$-9.28583$&$19.31$&$18.56$&$-63.3 \pm0.4 $&$4898_{-2  }^{+1  }$&$0.5_{-0.1}^{+0.0}$&$-2.0_{-0.0}^{+0.0}$&8.43&Y\\
        RGB9&$338.43417$&$-9.23303$&$20.99$&$20.28$&$-212.3\pm3.1 $&$5219_{-147}^{+161}$&$4.8_{-0.7}^{+0.5}$&$-1.0_{-0.2}^{+0.4}$&2.33&N\\
        BHB1&$338.49192$&$-9.43825$&$20.76$&$20.84$&$-70.6 \pm4.1 $&$7825_{-127}^{+132}$&$3.8_{-0.2}^{+0.2}$&$-0.9_{-0.2}^{+0.3}$&1.74&Y\\
        BHB2&$338.49288$&$-9.40789$&$20.65$&$20.71$&$-65.2 \pm3.6 $&$7780_{-54 }^{+97 }$&$3.5_{-0.1}^{+0.0}$&$-1.5_{-0.3}^{+0.4}$&1.86&Y\\
        BHB3&$338.49775$&$-9.41033$&$21.68$&$21.74$&$201.9 \pm11.9$&$8407_{-192}^{+193}$&$5.8_{-0.3}^{+0.2}$&$-0.2_{-0.4}^{+0.4}$&1.9 &N\\
        BHB4&$338.49167$&$-9.32186$&$20.6 $&$20.51$&$-62.6 \pm4.4 $&$7578_{-38 }^{+17 }$&$3.5_{-0.1}^{+0.0}$&$-1.9_{-0.3}^{+0.3}$&1.98&Y\\
        BHB5&$338.48112$&$-9.29314$&$20.56$&$20.62$&$-71.5 \pm4.5 $&$7788_{-144}^{+164}$&$3.2_{-0.2}^{+0.2}$&$-0.6_{-0.3}^{+0.1}$&1.92&Y\\
        BHB6&$338.50117$&$-9.308  $&$21.27$&$21.41$&$-94.9 \pm16.9$&$8216_{-227}^{+191}$&$3.9_{-0.1}^{+0.1}$&$-3.1_{-0.9}^{+1.1}$&1.55&Y\\
        \hline
    \end{tabular}
\end{table*}
\begin{figure*}
    \includegraphics[width=\textwidth]{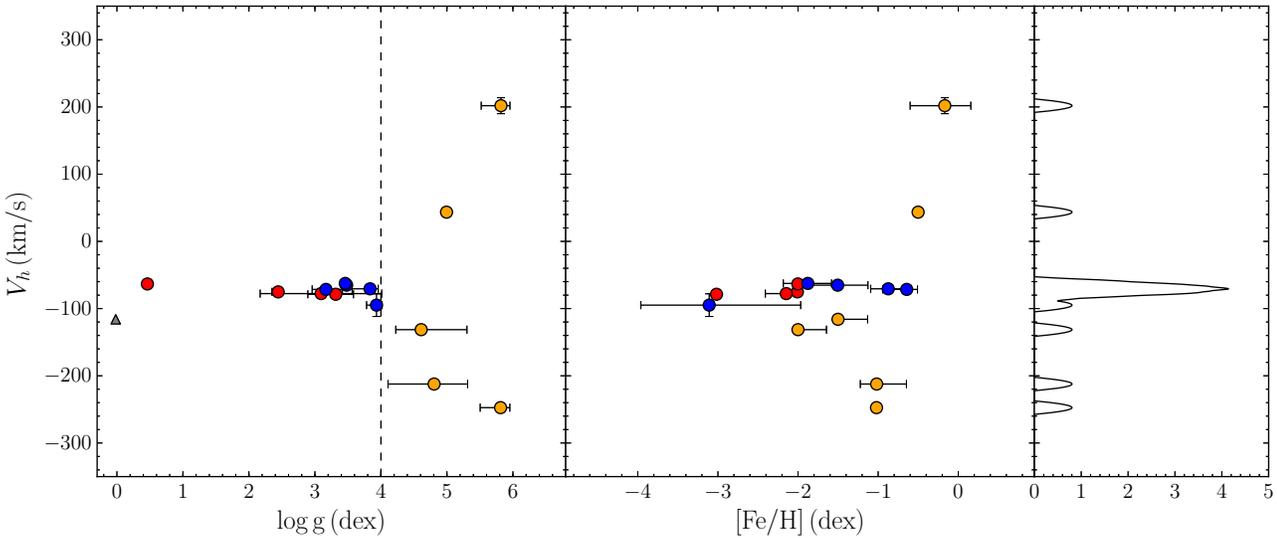}
    \caption{Heliocentric radial velocity plotted against $\log{\rm{g}}$
    (left) and $[$Fe/H$]$ (middle) for 15 stars with observed spectra. 
    Stars in orange have $\log{\rm{g}}>4$, and are likely field
    contamination.  The gray dot is the star without a good $\log{\rm{g}}$
    measurement.  Rightmost panel shows the generalized histogram for the
    radial velocity with a Epanechnikov kernel with 10
    km\,s$^{-1}$ bandwidth.  The velocity distribution of likely 
	members has a very well defined peak at -71 km\,s$^{-1}$
} \label{fig:Specdist}

\end{figure*}

\begin{figure}
    \includegraphics[width=0.98\columnwidth]{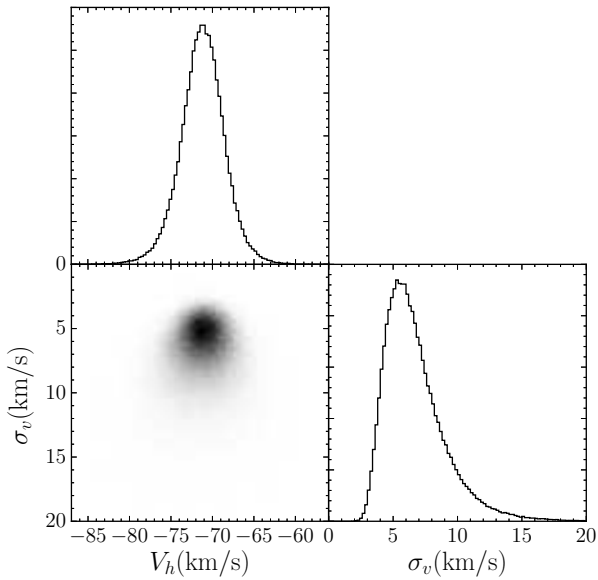}
    \caption{Posterior distribution for the heliocentric velocity and
      velocity dispersion of Aqu~2.  These are found by modelling
      the observed velocity distribution by a single Gaussian. Both
      posterior distributions shows a well defined maxima, with
      $\sigma_V$ having a long tail towards larger dispersions.
    } \label{fig:Specposterior}
\end{figure}

Figure \ref{fig:Specdist} displays the distribution of heliocentric
velocities as a function of surface gravity $\log{\rm{g}}$ (left
panel) and metallicity $[$Fe/H$]$ (middle panel) for all stars in the
spectroscopic dataset. We conjecture that stars with $\log{\rm{g}}>4$
are likely foreground contaminants, i.e. nearby MS stars. This
hypothesis is supported by a large spread in velocity observed for
this subgroup. In contrast, the velocity distribution of 9 stars (4
RGBs and 5 BHBs) with $\log{\rm{g}}<4$ has a well-defined narrow peak
at V$_h$$\sim$ -71\,km\,s$^{-1}$. This confirms that Aqu-2 is indeed a
{\it bona-fide} Milky Way satellite. It is also important that all the
BHB candidate stars have velocities consistent with the peak.
We measure the systemic velocity $V_h$ and the velocity dispersion
$\sigma_V$ of Aqu~2, by modelling the velocity distribution of likely
star members, i.e. the stars with $\log{\rm{g}}<4$ as a single
Gaussian with mean $V_h$ and dispersion $\sigma_v$. We sample the likelihood of the model using {\it emcee}
assuming flat priors for the mean and the Jeffreys priors for the
dispersion. The satellite mean velocity is thus determined to be
$V_h=-71.1 \pm 2.5$ km\,s$^{-1}$, corresponding to $V_{\rm GSR}=49$
km\,s$^{-1}$ assuming the Local Standard of Rest of 235 km\,s$^{-1}$
and the Solar peculiar motion as measured by \citet{uvw}. Aqu~2's
velocity dispersion is $\sigma_V=5.4^{+3.4}_{-0.9}$ km\,s$^{-1}$. Figure
\ref{fig:Specposterior} shows the corresponding posterior
distributions.

We can also estimate the mean metallicity of Aqu~2 by using the 4 RGB
member stars: $[$Fe/H$]=-2.3\pm0.5$. Notice a quite large error-bar
that is expected given the small number of stars, low resolution of
the spectra and step-size of 0.5\,dex in the template grid used. 
  However, most importantly, the spectroscopic metallicity measured is
  consistent on the 1 $\sigma$ level with the metallicity of the
  isochrone used in our photometric analysis.

According to the results of the previous Section, the extended size of 
Aqu~2 suggests that it should be classified as a dwarf galaxy, as
currently, there are no known star clusters that large. We further
firm up this classification by calculating the dynamical
mass. Using the estimator of \citet{Walker2009}, we calculate the mass
enclosed within the half-light radius to be
$2.7^{+6.6}_{-0.5}\times10^6 M_\odot$ which corresponds to a
mass-to-light ratio of $1330^{+3242}_{-227}$. This is the second
overwhelming piece of evidence that Aqu~2 is indeed a typical
dark matter dominated UFD. The following cautionary notes are worth
bearing in mind.  First, while the current data allow us to
resolve the velocity dispersion of Aqu~2, the value is inferred,
using only a small number of tracers. Furthermore, the
mass estimate may be affected by the strong elongation  of the dwarf, which
can be both a sign of an aspherical matter distribution as well as
out-of-equilibrium state of its member stars.

\section{Discussion and Conclusions}\label{sec:CONCL}

\begin{figure}
    \includegraphics[width=0.98\columnwidth]{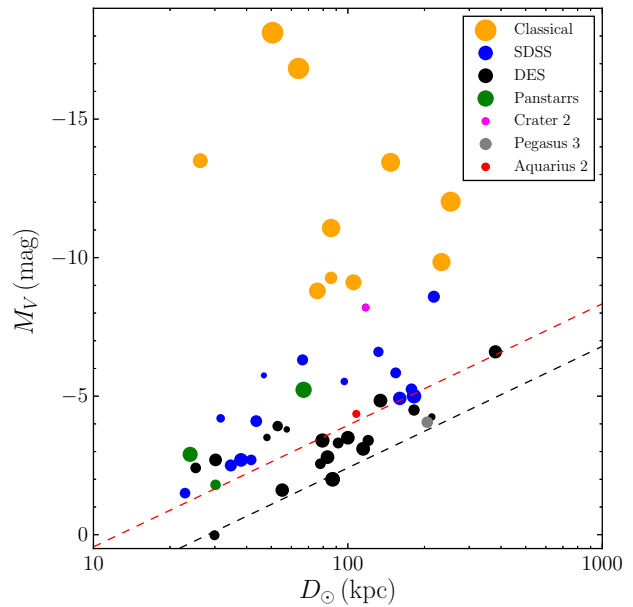}
    \caption{Absolute magnitude versus heliocentric distance for dwarf
      galaxies within 400\,kpc. The size of the markers reflects the
      surface brightness of the dwarfs, with the largest points
      showing the brightest galaxies.  We show classical dwarfs in
      orange, dwarfs discovered in SDSS in blue, dwarfs discovered in
      DES in black, and dwarfs discovered in PanStarss in green.  The
      red/magenta/gray points shows the position of
      Aqu~2/Crater~2/Pegasus~3. The red dashed line shows the
      detectability limit for SDSS as determined by
      \citep{Koposov2008} and the black dashed line the approximate
      limit for DES \citep{Jethwa2016}.  } \label{fig:MagVdist}
\end{figure}

\begin{figure*}
    \includegraphics[width=\textwidth]{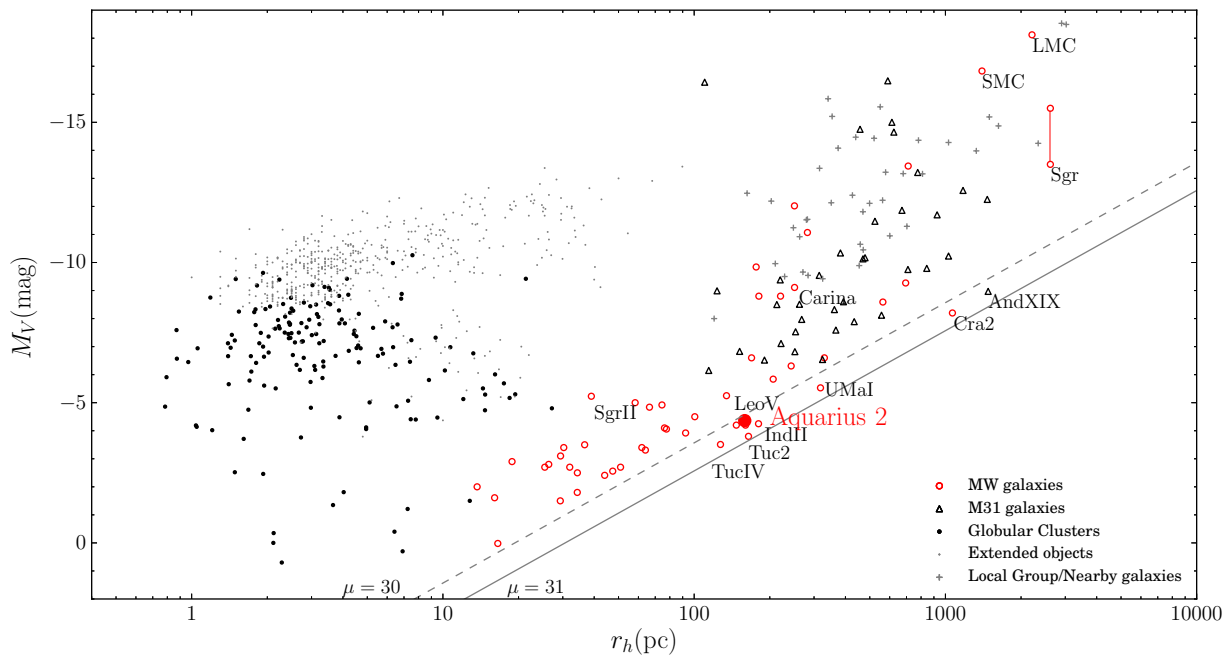}
	\caption{Absolute magnitude versus half-light radius.  Local
          galaxies from \citet{McConnachie2012} (updated September
          2015) are shown with different symbols.  Dwarf galaxy
          satellites of the Milky Way are shown with red open circles,
          the M31 dwarfs with black unfilled triangles, and other
          nearby galaxies with gray crosses. Note that for the Sgr
          dwarf, we also give the estimate of the progenitor's
          original luminosity from \citet[][]{MNO2010}. Black dots are
          the Milky Way globular clusters measurements from
          \citep{Harris2010,Belokurov2010,Munoz2012,Balbinot2013,Kim2015a,Kim2015b,Kim2015c,Laevens2015,Weisz2015}
          and gray dots are extended objects smaller than 100\,pc from
          \citep{Brodie2011}. The black solid (dashed) line
          corresponds to the constant level surface brightness within
          half-light radius of $\mu=31$ (30)
          $\rm{mag}\,\rm{arcsec}^{-2}$, which is approximately the
          surface brightness limit of the searches for resolved
          stellar systems in the SDSS \citep{Koposov2008}.  The
          position of the recently discovered Crater~2
          \citep{Torrealba2016} is also shown.} \label{fig:SizeLum}
\end{figure*}

Figure~\ref{fig:MagVdist} shows the distribution of the Milky Way
dwarf satellites in the plane of absolute magnitude $M_V$ and
heliocentric distance $D_{\odot}$. Objects are coloured according to
the imaging survey they were discovered in. The so-called Classical
dwarfs (orange) appear to be detectable at any distance throughout the
virial volume of the Galaxy, independent of their absolute
magnitude. This is unsurprising given their high intrinsic
luminosities. However, the UFDs, nominally identified as dwarfs with
$M_V>-8$ show a strong selection bias as a function of heliocentric
distance. According to \citet{Koposov2008}, the detectability of a
satellite as traced by its stellar members, depends primarily on
its surface brightness and heliocentric distance. So far, most of the
dwarfs discovered are brighter than $\sim 30-31$ mag\,arcsec$^{-2}$ as
illustrated in Figure~\ref{fig:SizeLum}. Curiously, all satellites
with $r_h<100$\,pc, lie above the $\mu=30$ mag arcsec$^{-2}$
line. However, the larger objects seem to obey a slightly fainter
detection limit, namely $\mu = 31$ mag arcsec$^{-2}$. To help analyze
the distribution of the satellites in the 3D space spanned by their
luminosity, surface brightness and distance, the symbol size in
Figure~\ref{fig:MagVdist} represents the effective surface
brightness of objects, with bigger circles corresponding to smaller
(i.e. brighter) values of $\mu$.

As discussed in \citet{Koposov2008}, the distance of a satellite 
plays an important role in determining whether it is going to be detected or
not. At a given heliocentric distance, of all objects below the
nominal surface brightness limit (as discussed above), only those
brighter than a certain limiting absolute magnitude can be securely
identified. \citet{Koposov2008} provide an expression for the limiting
absolute magnitude as a function of distance. This model is shown as a
dashed red line in Figure~\ref{fig:MagVdist}. Indeed, all SDSS-based
detections appear to lie above this detection boundary. Also shown
(dashed black line) is a version of the detectability limit derived by
\citet{Jethwa2016} for the objects identified in the DES data. Amongst
the SDSS objects, together with Leo~IV, Leo~V and Her, Aqu~2 forms a group
that sits nearest to the nominal luminosity boundary. Moreover, it is
also one of the lowest surface brightness satellites beyond 100 kpc
from the Sun. Thus, Aqu~2 is the only dwarf galaxy detected at both
the luminosity and the surface brightness limits. This is probably the
reason why this object had not been identified before.

The key reason for the surface brightness detection limit for
Milky Way satellites as measured by \citet{Koposov2008} is the balance
between the foreground density of stellar contaminants, the background
density of misclassified compact galaxies and the number of
satellite member stars. However, when BHB stars are used for
satellite detection, they represent a completely different regime, as
the surface density of contaminating foreground A-type stars and background
compact galaxies are orders
of magnitude lower compared to using MSTO or RGB-colored
stars. Furthermore, for the horizontal branch to be populated, the
luminosity of the satellite needs to be more than a certain critical
value. Thus, it would not be surprising if the actual combined BHB+RGB
detection boundary deviated somewhat from the simple limiting surface
brightness prescription of \citet{Koposov2008}.

Moreover, Figure~\ref{fig:SizeLum} hints at an intriguing possibility
that the intrinsic distribution of the Milky Way satellites forms a
reasonably tight sequence in luminosity-size space along the line
connecting the LMC and the newly discovered Aqu~2. The one object that
clearly appears to lie off this sequence is the Sgr dwarf, which is
known to have lost most of its stars \citep[see e.g.][]{MNO2010},
entering the last throws of the tidal disruption.  Objects like
UMa~I, Leo~V, Ind~II would then represent the more frequent occupants of
this dwarf-galaxy main sequence, while currently observed objects with
smaller sizes ($r_h<100$\,pc) may either represent a sub-population of
tidally disrupting objects, or those with properties transitional
between globular clusters and dwarf galaxies.

\subsection{The LMC connection?}

As illustrated in Figure~\ref{fig:MagVdist}, between $\sim$60 and
$\sim$130 kpc, there appears to be a paucity of SDSS UFDs detected,
with only one such object known, namely Bootes I. Aqu~2, at
$\sim$110 kpc, helps to fill in this gap, but is almost $\sim$2
magnitudes fainter. Of course, there are plenty even fainter dwarf
satellites in this heliocentric distance range, i.e. those discovered
in the DES data. However, these cluster spatially around the LMC and the SMC,
and, as shown by \citet{Jethwa2016}, are likely to be associated with
the accretion of the Clouds by the Galaxy.

Aqu~2 lies just 9 kpc from the current orbital plane of the LMC,
as defined by the most recent measurement of the LMC 3D velocity
vector \citet{Kallivayalil2013}.  We now consider whether this is a
chance alignment, or signals some association between these Galactic
satellites.  This question is further motivated by the recent
discovery of 17 dwarf galaxies and dwarf galaxy candidates discovered
in data taken from the Dark Energy Survey
\citep[DES][]{Koposov2015,Bechtol2015,Hor2,DrlicaWagner2015} many of
which have been shown to be likely associated with the Magellanic
Clouds \citep{Yozin2015,Deason2015,Jethwa2016}, thus confirming a generic
prediction of cold dark matter cosmological models whereby the
hierarchical build up of structure leads to groupings of dwarf
satellite galaxies in the Galactic halo.  Most recently,
\citet{Jethwa2016} investigated this by building a dynamical model of
satellites of the Magellanic Clouds.  This included the gravitational
force of the Milky Way, LMC and the SMC, the effect of dynamical friction on
the orbital history of the Magellanic Clouds, and marginalized over
uncertainties in their kinematics.  Fitted against DES data, their
results suggested that the LMC may have contributed $\sim70$ dwarfs to
the inventory of satellite galaxies, i.e. $\sim$30\% of the total
population.

\begin{figure}
    \includegraphics[width=0.98\columnwidth]{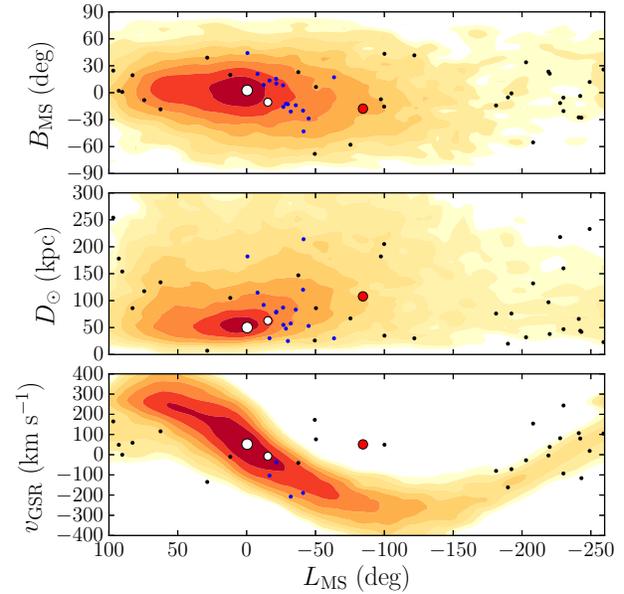}
    \caption{Comparison of the phase space distribution of known dwarf
      galaxies within 300\,kpc (plotted symbols) with a simulated
      distribution of LMC satellites (coloured contours).  The top
      panel shows the on-sky distribution in Magellanic Stream
      co-ordinates $(L_\mathrm{MS},B_\mathrm{MS})$.  The middle panel
      shows heliocentric distance, and the bottom panel line-of-sight
      velocity in the Galactic standard of rest, both as functions of
      $L_\mathrm{MS}$.  Contour colours step in factors of two
      density, normalised independently per panel.  The LMC and SMC
      are represented by the small and large white circles, other
      dwarf galaxies as dots (blue if discovered in DES, black
      otherwise) and Aqu~2 as the red symbol.  
		}
    \label{fig:LMC_sats}
\end{figure}

Figure~\ref{fig:LMC_sats} shows the distribution of LMC satellites
from the maximum likelihood model of \citet{Jethwa2016} as coloured
contours, with known dwarfs shown as plotted symbols.  The top panel
shows the on-sky distribution in Magellanic Stream co-ordinates
$(L_\mathrm{MS},B_\mathrm{MS})$ \citep[defined in][]{Nidever2008}, where
latitude $B_\mathrm{MS}=0$ traces the centre of the HI gas
distribution, which is slightly offset from the LMC proper motion
vector.  We see that Aqu~2 (shown by the red symbol) lies at
relatively small Magellanic latitude $B_\mathrm{MS}=-17.8$ deg, and -
as shown in the middle panel - is well within the scope of the
simulated distance distribution of Magellanic satellites.  The bottom
panel, however, shows that at its Magellanic longitude, Aqu~2 is
moving too slowly to be a member of the trailing debris of LMC
satellites.  This is generically true over the grid of models presented by
\citet{Jethwa2016} and hence we disfavour a Magellanic origin for
Aqu~2, preferring instead a scenario where it is part of a
virialised population of Milky Way satellites.

\subsection{Summary}

Aqu~2 was discovered at the very edge of the VST ATLAS footprint as an
overdensity of RGB stars. What made it stand out amongst the several
other candidates with similar significance is the presence of a
prominent BHB. Moreover, the object also appeared in the list of
significant overdensities produced using the SDSS DR9 data, located,
equally, not very far from the edge of the footprint. To establish the
nature of the candidate, we have obtained deep follow-up imaging of
Aqu~2 with Baade's IMACS camera, as well as spectroscopy with Keck's
DEIMOS, which fully confirmed that the object is a true Milky Way
satellite. Given the satellite size of $\sim160$\,pc and the stellar
velocity dispersion of $5.4^{+3.4}_{-0.9}$ km\,s$^{-1}$, Aqu~2 is
undoubtedly a dwarf galaxy. The shape of Aqu~2 is mildly elliptical
$1-b/a\sim 0.4$, and the radial velocity indicates that Aqu~2 is
currently moving away from the Galacic center, on its way to
apo-galacticon. Taken at face value, its structural and kinematic
parameters imply an extremely high mass-to-light ration of
$M/L=1330^{+3242}_{-227}$. We stress, however, that the determination
of the structural parameters of Aqu~2, particularly the half-light
radius, can be significanlty improved with wider field-of-view
observations. Similarly, the velocity dispersion measurement would
benefit from a larger sample of satellite member stars.  The discovery
of Aqu~2 is a testimony to the fact that many more faint dwarfs are
surely still lurking in the SDSS and VST ATLAS datasets.

\section*{Acknowledgments} 

Support for G.T. is provided by CONICYT Chile. The research leading to
these results has received funding from the European Research Council
under the European Union's Seventh Framework Programme
(FP/2007-2013)/ERC Grant Agreement no.  308024. This research was made
possible through the use of the AAVSO Photometric All-Sky Survey
(APASS), funded by the Robert Martin Ayers Sciences Fund.

\bibliographystyle{mn2e}
\bibliography{biblio}

\label{lastpage}

\end{document}